\documentclass[a4paper]{article}

\usepackage{INTERSPEECH2022}

\usepackage{enumitem}
\usepackage{slantsc}
\usepackage{multirow}
\usepackage{dsfont}
\usepackage{color}
\usepackage{arydshln}
\usepackage{booktabs}
\usepackage[tracking=true]{microtype}
\usepackage{cite}
\usepackage{xcolor}
\usepackage{hyperref}
\usepackage{xspace}

\title{\vspace{-2ex} SAQAM: Spatial Audio Quality Assessment Metric}

\name{\scalebox{.9}[1.0]{Pranay Manocha$^{1*}$\thanks{$^{*}$Work done during internship at Meta.},
      Anurag Kumar,$^{2}$
      Buye Xu,$^{2}$
      Anjali Menon,$^{2}$
      Israel D. Gebru,$^{2}$
      Vamsi K. Ithapu,$^{2}$
      Paul Calamia$^{2}$ \vspace{-0.08in}}}
\address{
  $^1$Department of Computer Science, Princeton University, Princeton, NJ, USA\\
  $^2$Meta Reality Labs Research, Redmond, WA, USA}
  
\email{pmanocha@cs.princeton.edu, \{anuragkr90, xub, aimenon, idgebru, ithapu, pcalamia\}@fb.com
\vspace{-1ex}
}

\DeclareUnicodeCharacter{2212}{-}
\newcommand{\ignorethis } [1] {}

\newcommand{\etal       }     {{et~al.}}

\newcommand{\eg         }     {{e.g.}}

\newcommand{\Reals      }     {{\textrm{I\kern-0.18em R}}}

\renewcommand{\vec      } [1] {{\text{\boldmath $\mathbit{#1}$}}}

\newcommand{\change     } [1] {\mbox{{\footnotesize $\Delta$} \kern-3pt}#1}

\newcommand{\acronym}[1] {{\uppercase{#1}}}

\newcommand{\PESQ}   {\acronym{Pesq}}

\newcommand{\AMBIQUAL}   {\acronym{Ambiqual}}
\newcommand{\OURS}   {\acronym{SAQAM}}
\newcommand{\DPLM}   {\acronym{Dplm}}
\newcommand{\BAMQ}   {\acronym{Bamq}}

\newcommand{\listeningquality}   {{{LQ}}}
\newcommand{\spatializationquality}   {{{SQ}}}
\newcommand{\overallquality}   {{{OVRL}}}

\makeatletter
\newcommand*{\etc}{%
    \@ifnextchar{.}%
        {etc}%
        {etc.\@\xspace}%
}
\makeatother

\begin{document}

\maketitle
\begin{abstract}
Audio quality assessment is critical for assessing the perceptual realism of sounds.
However, the time and expense of obtaining ``gold standard'' human judgments limit the availability of such data.
%
%
%
For AR\&VR, good perceived sound quality and localizability of sources are among the key elements to ensure complete immersion of the user.
Our work introduces \OURS\, which uses a multi-task learning framework to assess listening quality (\listeningquality) and spatialization quality (\spatializationquality) between \emph{any} given pair of binaural signals without using any subjective data.
We model \listeningquality\ by training on a simulated dataset of triplet human judgments, and \spatializationquality\ by utilizing activation-level distances from networks trained for direction of arrival (DOA) estimation.
%
%
%
%
We show that \OURS\ correlates well with human responses across four diverse datasets.
Since it is a deep network, the metric is differentiable, making it suitable as a loss function for other tasks. 
For example, simply replacing an existing loss with our metric yields improvement in a speech-enhancement network.

\end{abstract}

%


\section{Introduction}

%
Audio quality plays a fundamental role in many applications affecting the quality of listening experiences like AR\&VR.
To create an immersive AR\&VR experience, both graphics and spatial audio need to be of high quality and synchronized to the user’s head movements in real time. 
In addition, audio sources must be accurately positioned such that they are localized at the targeted location. 
It is well-known that a mismatch between the stored auditory schemata of the expected listening scene and the perceived schemata of a simulation yields a significant decrease in presence and immersion in these multi-sensory systems~\cite{werner2014influence}.
Therefore, sound-quality evaluation tests are critical since they provide the necessary user feedback that drives improvements in these technologies.
Listening tests in AR\&VR via user studies demand more time and effort than conventional audio tasks as they require assessment across many attributes~\cite{lindau2014spatial} (\eg\ listening quality, spatial localization \etc). 
Thus, automatic (objective) SQA methods are more practical.

Recent research has focused on learning a spatial audio metric from known binaural auditory cues like Interaural Level Differences (ILD), Interaural Time Differences (ITD) and Cross-Correlation (IACC) between signals entering the left and the right ear~\cite{kampf2010standardization,flessner2017assessment,seo2013perceptual,takanen2012binaural}. However, these methods suffer from various general drawbacks. First, these methods are designed to only assess \emph{spatialization quality (\spatializationquality}) that ensures how accurately the test sources are positioned as compared to the reference, but cannot assess effects of audio fidelity degradation's, also referred to as \emph{listening quality (\listeningquality}). 
It includes undesired sounds and distortions that add artifacts to the audio signal and cause a loss in immersiveness.
%
Second, these methods are full-reference and always require a clean signal for comparison (that share the same `content' - \emph{matched reference}), and thus cannot be applied in scenarios where the paired clean reference is unavailable.
Third, they work well under anechoic conditions but are inaccurate under reverberant conditions.
%
%
%
Fourth, they assume that the test signal is created in relation to the reference, meaning that the two signals are time-sligned and of equal length which may not be valid.
%
Finally, none of the metrics are differentiable, and thus cannot be directly leveraged as a training objective in downstream tasks.
Addressing a few concerns, researchers have proposed \AMBIQUAL~\cite{narbutt2020ambiqual} that compares multi-channel ambisonic recordings across \listeningquality\ and \spatializationquality. However, it does not account for head listening direction, and the influence of Head Related Transfer Functions (HRTFs) in the resulting recording. Moreover, it also requires access to a clean, matched reference, and is also non-differentiable.

Manocha \etal~\cite{manocha2021dplm} proposed a full-reference deep perceptual spatialization metric (\DPLM) that evaluates the similarity of binaural presentation in terms of localization under realistic, echoic conditions. \DPLM\ computes deep-feature distances between the full-feature activation stacks of direction-of-arrival (DOA) models to assess \spatializationquality\, and correlates well with human perceptual judgments. Moreover, the metric is differentiable and does not demand access to a clean reference or time-aligned signals. However, it does not assess \listeningquality. 



In this paper, we introduce a novel metric that addresses some of the above concerns. 
We propose \OURS\ that evaluates the \listeningquality\ and \spatializationquality\ between \emph{any} pair of binaural signals. The inspiration for the approach comes from human’s ability to do the same. Given two completely random audio (``non-matched'') recordings, it is highly likely that a human would be able to compare them across quality attributes irrespective of the audio content~\cite{manocha2021noresqa}.
We first design a \listeningquality\ assessment model that assesses the audio fidelity degradations, and is trained on a carefully simulated dataset of perturbations that mimic realistic environments. Next, these recordings are presented to the model as triplets, and the model is trained using triplet metric learning~\cite{wang2014learning,hoffer2015deep}, by carefully selecting triplets to have different audio content(s) which ensures content invariance.
Next, to assess \spatializationquality, we build on top of the \DPLM~\cite{manocha2021dplm} approach that addressed the perceptual characteristics of localizing sounds. 
We combine these two tasks using a multi-task learning framework~\cite{caruana1997multitask} to leverage the useful information contained in related tasks to improve generalization performance on both tasks.
%
%
We show that, even in the absence of explicit perceptual training, these distances correlate well with human perceptual judgments (both via objective and subjective evaluations). We also show that the resulting metric generalizes well even for distinct (yet related) tasks such as audio codecs, and binaural reproduction from mono or multi-channel signals. Finally, we show that adding \OURS\ to the loss function yields improvements to the existing state-of-the-art model for binaural speech enhancement.
%
%

%



\section{The SAQAM Framework}

\begin{figure}[t!]
\vspace{-4\baselineskip}
\centering
\setlength{\tabcolsep}{4pt}
\includegraphics[width=\columnwidth]{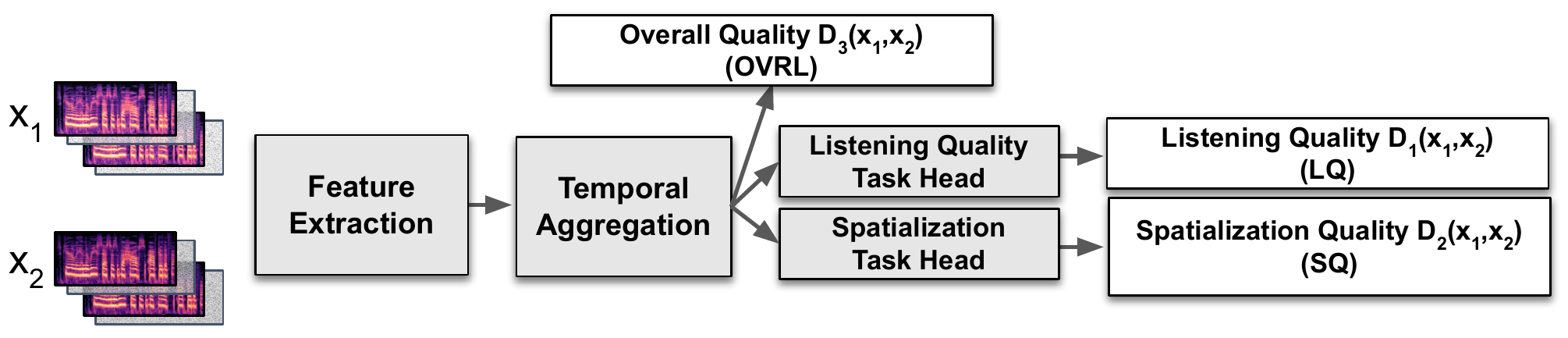}\caption{\textbf{\OURS\ framework}: Our framework takes two inputs $x_1$ and $x_2$, and estimates a rating for each attribute - spatialization quality (\spatializationquality), listening quality (\listeningquality) as well as overall quality that combines all above attributes (\overallquality). Note that all distances are deep-feature distances. }

%
\vspace{-0.1in}
\label{framework}
\end{figure}

Our framework \OURS\, is designed to assess the quality of a given pair of binaural signals across various attributes. In this paper, we consider two such attributes: listening quality (LQ) and spatialization quality (SQ), but the framework can be easily extended to include more such attributes.
Additionally, the \OURS\ framework can also estimate a combined, overall attribute (\overallquality) that implicitly combines all individual attributes.
Refer to Fig~\ref{framework}. Given two binaural signals ($x_\mathtt{1}$ and $x_\mathtt{2}$), our goal is to compute distance functions $D_1(x_\mathtt{1},x_\mathtt{2})$, $D_2(x_\mathtt{1},x_\mathtt{2})$ and $D_3(x_\mathtt{1},x_\mathtt{2})$ that characterize \listeningquality, \spatializationquality, and \overallquality\ between the signals. The distance function is designed to be \emph{non-negative} and \emph{monotonic}, thereby making it a pseudo-metric (we do not impose triangle or associative properties).

\subsection{Multi-task learning}
\label{comprehensive_model}

Multi-task learning (MTL)~\cite{caruana1997multitask} has been beneficial to many speech applications~\cite{fu2016snr,chen2015multitask}. MTL aims to leverage useful information contained in multiple related tasks to help improve the generalization performance on all tasks. In our case, we consider \listeningquality\ and \spatializationquality\ prediction as the two tasks. MTL makes the model easier to use - in the sense that if only the spatialization results are needed or if only the listening quality is required, the appropriate task-head can be used while discarding the others.
Moreover, since MTL encourages learning shared information, the common shared layers of the model can be used in cases where an overall estimate of quality (\overallquality) is required rather than individual estimates of \listeningquality\ and \spatializationquality.

\noindent {\bf \listeningquality\ head:} is trained using a triplet comparison dataset where we simulate human judgments on a simple question: ``Is A or B closer to reference C?”. We algorithmically modify clean recordings under various real-life degradations commonly found in audio processing tasks including compression, additive noise, speech distortions and binaural recorded noises across sources.

\noindent {\bf \spatializationquality\ head:} is re-purposed from training a sound source-localization model that predicts the direction-of-arrival (DOA) of a given sound source, similar to \DPLM~\cite{manocha2021dplm}. We train DOA models with carefully designed input perturbations as data augmentations that mimic realistic environments.


\subsection{Architecture}
\label{architecture}
Refer to Fig~\ref{architecture}. The architecture of both tasks consists of 3 components: feature-extraction block, temporal aggregation block, and task specific heads for \listeningquality\ and \spatializationquality.
%
In the MTL framework, the first two blocks are shared between tasks.


For the feature-extraction block, we used an Inception~\cite{szegedy2015going} styled architecture. The 6-block
Inception network consists of 1x1, 3x3
and 5x5 filters, leading to 3x3 max-pooling, and a 1x2 maxpool
along the frequency dimension. For the temporal aggregation block, we use Temporal Convolutional Networks (TCNs) consisting of 4 temporal blocks., with each block consisting of 2 convolutional layers with a kernel size of 1x3.
We use weight normalization~\cite{salimans2016weight} along with dilated convolutions to increase the effective receptive field of our model.

\begin{figure}[t!]
\vspace{-4\baselineskip}
\centering
\setlength{\tabcolsep}{4pt}
\includegraphics[width=1\columnwidth]{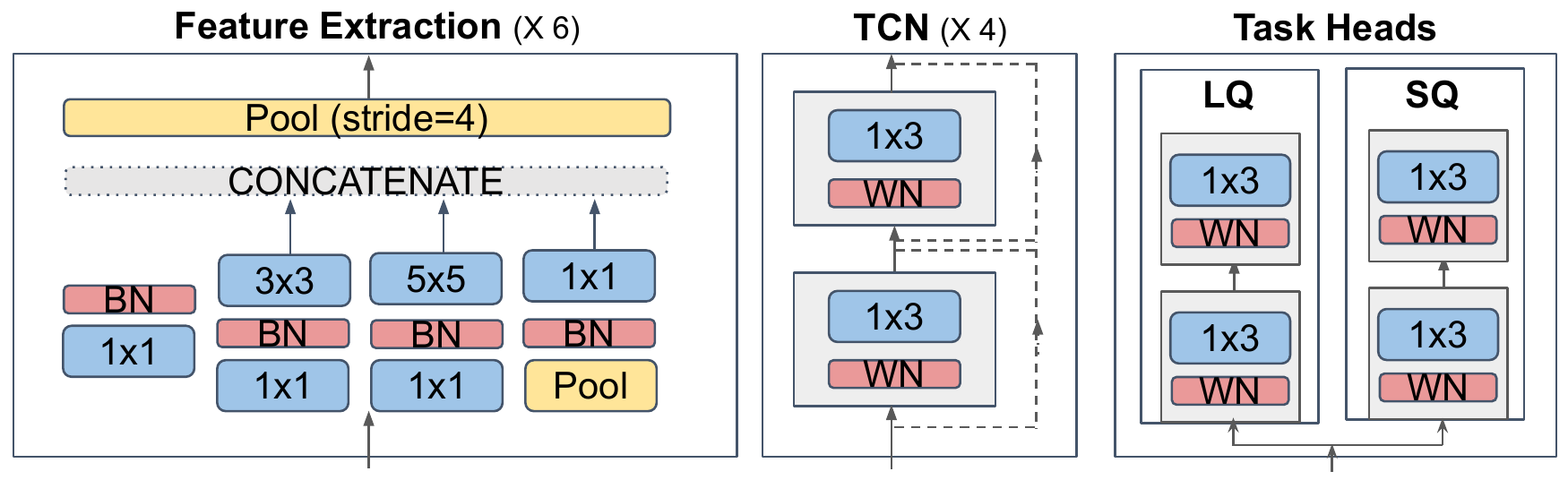}
\vspace{-4ex}
\caption{\textbf{\OURS\ architecture}: consists of feature extraction (Inception~\cite{szegedy2015going}), temporal aggregation (TCN), and multi-task learning heads, one each for \listeningquality\ and \spatializationquality. }

\vspace{-0.2in}
\label{architecture}
\end{figure}

For the \listeningquality\ task, we use 2 conv. layers to transform the embedding size to 64 per frame. These are then mean-pooled in the temporal dimension, the resulting embedding representing the \listeningquality\ of the signal. For the \spatializationquality\ task,  we divide azimuth ($\in(-180^{\circ}, 180^{\circ})$) into $50$ equally spaced bins, and the elevation ($\in(-90^{\circ}, 90^{\circ})$) into $25$ equal bins.
The resulting embedding from the aggregation stage is then mapped to a one-hot vector, representing the location of the sound source.
Note that both task-heads output a framewise estimate.

\subsection{Loss Functions}
\label{loss_functions}

We use different losses to train the \listeningquality\ and \spatializationquality\ heads. 
For \listeningquality, we use deep metric learning with a \emph{triplet-loss} as the basis to train the \listeningquality\ head. 
Formally, we define the triplets as a set $T = \{t_{i}\}_{i=1}^{N}$ where each triplet $t_i = \{x_a^i,x_p^i,x_n^i\}$ with $x_a$ the anchor sample, $x_p$ the positive sample, and $x_n$ the negative sample. Then, we define the triplet loss as:
\begin{equation}
L(t) = max\{ 0,D_1(x_a,x_p) - D_1(x_a,x_n) +\delta \}.
\label{eqn:triplet}
\end{equation}

\noindent Here $\delta$ is the margin value to prevent trivial solutions, and $D_1(x_1,x_2)$ is the deep-feature distance~\cite{manocha2020differentiable} between the full-feature activation stacks between the two audio inputs. 



For the \spatializationquality\ metric, we use the same \emph{classification} formulation as Manocha \etal~\cite{manocha2021dplm}. However, instead of the conventional cross-entropy loss, we use the earth mover's distance (EMD) as a loss function for training. It focuses on inter-class relationships, whereas plain cross-entropy loss ignores any relationship between classes.
This is especially useful if the classes are ordered since that leads to a closed form solution, which is:
\vspace{-0.08in}
\begin{equation}
 \mbox{\bf{EMD}}^{2}(\hat{p},p) = \sum_{n=1}^{N}(\hat{P_n} - P_n)^2
 \vspace{-0.05in}
\label{eqn:deepfeat}
\end{equation}
where $P_n$ is the $n$-th element of the cumulative density function of $p$, and $\hat{p}$ is the predicted output after softmax at the final layer. 
We follow a normal label distribution around the correct class, and use a discrete soft label distribution compared to one-hot:
 \vspace{-0.05in}
%
\begin{equation}
p_n = \begin{cases}
0.4 &\text{n=v}\\
0.2 &\text{n=v$\pm$1}\\
0.1 &\text{n=v$\pm$2}\\
0 &\text{otherwise}\\
\end{cases}
 \vspace{-0.05in}
\end{equation}
where $v$ is the index corresponding to the ground truth localization. Note that we also enforce continuity between the first and the last bins for the azimuth space.


\subsection{Usage}
%
%
Given a test input $x_1$, and a reference signal $x_2$, the \listeningquality\ score is obtained from the \listeningquality\ task head of the network by computing the deep-feature distances across activation maps $D_1(x_1,x_2)$. Similarly, the \spatializationquality\ score is obtained from the \spatializationquality\ task head of the network by computing the deep-feature distances across activation maps $D_2(x_1,x_2)$. Finally, the \overallquality\, score is obtained from from the shared body (temporal aggregation block) of the network by computing the deep-feature distances across activation maps $D_3(x_1,x_2)$. Note that the reference signal \emph{does not} need to contain the same speech (or content - hence \emph{`non-matching'}) as the test signal. 

\section{Experimental Setup}

\subsection{Datasets and training}
Speech recordings from the TIMIT~\cite{garofolo1993darpa} dataset are used as the
source for anechoic recordings ($\mathcal{D}_{\!c}$). We used a pool of 11 publicly available Binaural Room Impulse Response
(BRIR) databases including ADREAM~\cite{winter2016database}, AIR\_1\_4~\cite{jeub2009binaural}, BRAS~\cite{aspock2020bras}, Huddersfield~\cite{bacila2019360}, Ilmenau~\cite{mittag_christina_2016_206860}, IoSR~\cite{francombe2016iosr}, Oldenburg\_IE\_BTE~\cite{kayser2009database}, Rostock~\cite{erbes2015database}, TU Berlin~\cite{wierstorf2011free}, and Salford~\cite{satongar2014measurement}. The resulting pool contained approximately 125k BRIR pairs from 36 different rooms ($RT_{60}$ ranging from 0.08 to 0.97 sec). All single channel additive noises are sampled randomly from the DNS Challenge~\cite{reddy2020interspeech}.

For the perturbation set, we consider all degradations commonly found in various audio processing tasks including additive noise, speech distortions (\eg\ clipping and frequency masking, frequency resampling, pitch shifting), compression (e.g., mu-law and MP3), and recorded binaural sounds~\cite{foster2015chime,weisser2019ambisonic}. We also use the data collected using the binaural multichannel wiener filter (MWF)~\cite{cornelis2009theoretical} algorithm, and find that adding datasets and perturbations with subtle differences increases the robustness of our model to small differences.


To obtain the simulated triplet set ($t_i = \{x_a^i,x_p^i,x_n^i\}$), we first sample three clean recordings ($s_a,s_p,s_n$) from $\mathcal{D}_{\!c}$ and binauralize.
Next, a perturbation is sampled from the above mentioned perturbation set, and applied to ($s_a,s_p,s_n$) at three different levels uniformly sampled from a range \mbox{(-20dB to +30dB)} to create ($x_a,x_p,x_n$). It is ensured that the perturbation levels of $x_a$ and $x_p$ are closer than those for $x_a$ and $x_n$ to satisfy our triplet formulation.
Note here that the inputs to be model ($x_a,x_p,x_n$) are created by sampling \emph{different} clean recordings, but adding the \emph{same} noise at different levels.
This encourages the model to be invariant to recording content. These findings can be treated as a pilot study for future work on a dataset of subjective quality preferences.
%

We use an adaptive margin ($\delta$) strategy that starts out with a low value ($\delta=0.5$), and increases gradually ($\delta=1.50$) to ensure robust training~\cite{harwood2017smart}. 
%
We use 3-sec excerpts for training. For both heads, phase and magnitude spectrogram are extracted from the 2 channels using a 512-point DFT with a hamming window of length 512, and 50\% overlap.
We use a learning rate of $10^{−4}$ with a batch size of 64 for training. As part of online data augmentation to make the model invariant to small delay, we randomly add a 0.25s silence to the audio at the beginning or the end to provide shift-invariance property to the model. 
%
For the localization model, we follow the same training methodology as in Manocha \etal~\cite{manocha2021dplm}.

\begin{table}[b!]
\vspace{-3ex}
\centering
\setlength{\tabcolsep}{4pt}
\resizebox{1\columnwidth}{!}{
 \begin{tabular}{l l c c c c}
 \toprule
{\bf Name} & {\bf Attribute}  & {\bf CommonArea$\downarrow$} & \bf $MP_{10}\uparrow$ & \bf $MP_{25}\uparrow$ & {\bf Monotonicity$\uparrow$}  \\
 \toprule
\multirow{2}{*}{\bf Proposed}
&  {\bf SQ} &  0.19 &  0.92 &  0.89 &  0.94  \\ 
&  {\bf LQ} &  0.23 &  0.87  &  0.79 &  0.92    \\ 
\cdashline{1-6}
 \multirow{1}{*}{\bf \DPLM}
&  {\bf SQ} & 0.25 & 0.87 & 0.82 & 0.93  \\
 \bottomrule

\end{tabular}
}
\caption{\textbf{Objective evaluations}. Sec~\ref{obj_eval} describes
common area, mean precision, and monotonicity (using Spearman correlations) across simulated datasets.~$\uparrow$ or $\downarrow$ is better.}
\vspace{-6ex}
\label{objective_evaluation}
\end{table}

\subsection{Baselines}

For \listeningquality, since there does not exist any metric that assesses audio fidelity degradations of binaural signals, we compare it with a single channel quality metric like \PESQ~\cite{rix2001perceptual}. 
For \spatializationquality, we compare our approach to the \DPLM\ approach of Manocha \etal~\cite{manocha2021dplm}, and the \BAMQ\ approach of Flessner \etal~\cite{flessner2017assessment} that estimates the binaural cues at frame-level and combines them using a set of learned weights to output an overall quality between two recordings.

\section{Results}

\subsection{Objective evaluations}
\label{obj_eval}

Objectively, we compare various metrics on (i)~robustness
to content variations; (ii)~clustering in learned space; and (iii)~monotonic behavior with increasing perturbation levels. Refer to Table~\ref{objective_evaluation} for the results. Note that none of the above-mentioned baselines accepts non-matched inputs, except \DPLM, which only assesses \spatializationquality.

\begin{figure}[t!]
\vspace{-4ex}
\centering
\includegraphics[width=0.80\linewidth]{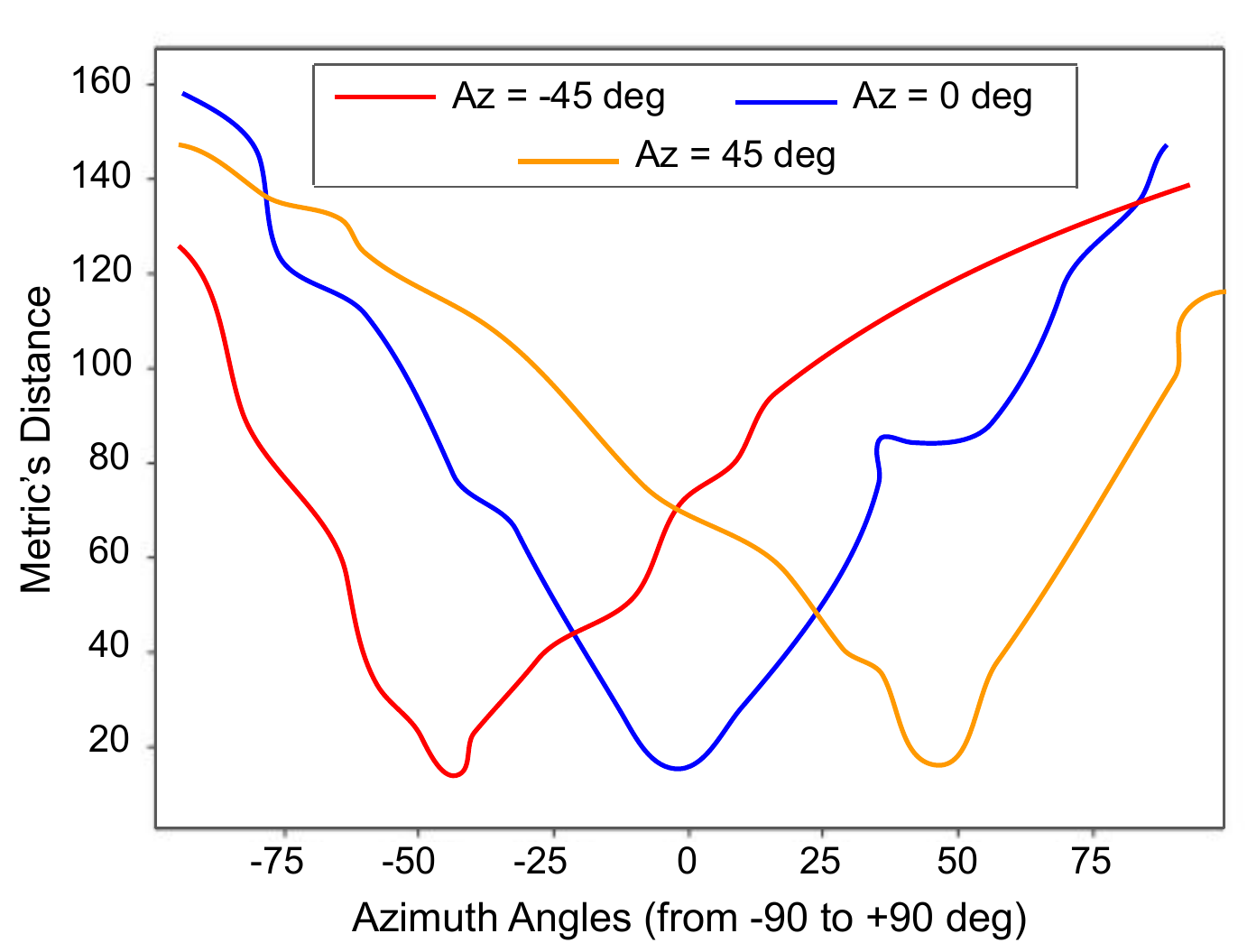}
\vspace{-2ex}
\caption{\textbf{\OURS\ with angular distance}: This shows \OURS's variation with angular distance for three fixed reference positions compared across \emph{non-matched} content}
\label{compare_static_dynamic}
\vspace{-4ex}
\end{figure}
\begin{table*}
\vspace{-6ex}
\centering
\renewcommand{\arraystretch}{1.1}
\resizebox{\textwidth}{!}{
 \begin{tabular}{l l c c c c c c c c c c c c c c c c c}
 \toprule
 \multirow{2}{*}{\bf Type} & \multirow{2}{*}{\bf Name} & \multicolumn{3}{c}{\bf P1 } & \multicolumn{2}{c}{\bf P1' }& \multicolumn{1}{c}{\bf P2}& 
 \multicolumn{3}{c}{\bf P3}& 
 \multicolumn{6}{c}{\bf P4} \\
 \cmidrule(lr){3-5} \cmidrule(lr){6-7} \cmidrule(lr){8-8} \cmidrule(lr){9-11} \cmidrule(lr){12-17}
 & &\bf Speech &\bf Castanets & \bf Guitar & \bf Speech & \bf Castanets & \bf Music & \bf Speech & \bf Pink Noise & \bf Guitar & \bf Pink Noise & \bf Vocals & \bf Castanets & \bf Glocken & \bf EM & \bf AM \\
 \cmidrule(lr){1-17}
 \multirow{1}{*}{\bf Conven.}
  & {\bf PESQ} & 0.01 & 0.23  & 0.12 & 0.07 & 0.17 & 0.01 & 0.09 & -0.20 & 0.08 & -0.02 & 0.11 & 0.13 & -0.05 & -0.04 & 0.08 \\
  \cdashline{1-17}
 \multirow{2}{*}{\bf Proposed}
  & {\bf SQ} & 0.96 &  0.95  &  \bf 0.94 &  0.88 &  1.0 &  0.52 &  0.78 & 0.23 & 0.26 &  0.53 & 0.60 &  0.49 &  \bf 0.47 &  0.69 & 0.82 \\
 & {\bf OVRL} & \bf 1.0 &  \bf 1.0  &  \bf 0.94 &  \bf 1.0 &  \bf 0.94 &  \bf 0.64 &  \bf 0.86 & 0.40 & \bf 0.69  & \bf 0.66 &  \bf 0.89 &  \bf 0.77 &  \bf 0.47 & \bf 0.70 & \bf 0.88 \\
  \cdashline{1-17}
 \multirow{2}{*}{\bf Baseline}
 & {\bf DPLM} &  0.94 &  0.94  &  \bf 0.94 &  0.83 &  \bf 0.94 &  0.45 &  0.69 & 0.22 & 0.06 &  0.53 & 0.61 &  0.42 &  \bf 0.47 &  0.67 &  0.83 \\
  & {\bf \BAMQ\,} & 0.03 & 0.83  & 0.09 & 0.52 & 0.77 & -0.17 & 0.42 & \bf 0.65 & 0.08 & -0.02 & 0.36 & 0.11 & -0.05 & 0.23 & 0.18 \\
 \bottomrule
\end{tabular}
}

\caption{\textbf{Subjective evaluation (1)}: Compared across overall ratings. Models include: Conventional model (e.g., \PESQ), our proposed models (\textit{SQ} and \textit{OVRL} models), Baseline models including \DPLM\ and \BAMQ. Spearman Correlation (SC). $\uparrow$ is better.}
\label{table_mos1}
\vspace{-2ex}
\end{table*}

\noindent {\bf Robust to content variations}    
To evaluate robustness to non-matched recordings, we create a test dataset of two groups: first group consists of pairs of recordings with the same quality attributes (\listeningquality\ and \spatializationquality) but different speech content; the other group consists of recordings with different quality attributes and speech content.
We calculate the common area between these normalized distributions. Our \OURS\ metric has the \emph{lowest} common area, suggesting that it is agnostic to speech content.

\noindent {\bf Clustered Embed. Space:}
To better understand the learnt representations, we evaluate the retrieval performance of our model using Precision@K to measure the quality of the top $K$ items in the ranked list. This is done for both \listeningquality\ and \spatializationquality\ task heads. We divide both attributes into 10 different perturbation-level groups, with each group consisting of 100 recordings with the same perturbation and level, but non-matched content. We take randomly selected queries and calculate the number of correct class instances in the top K retrievals.
We report the mean precision@K ($\text{MP}_{k}$) over all test queries which shows high precision retrievals that suggests that the embeddings indeed capture these quality attributes.



\begin{table}
\vspace{-0.20in}
\centering
\renewcommand{\arraystretch}{1.1}
\renewcommand{\arraystretch}{1.1}
\resizebox{\columnwidth}{!}{
 \begin{tabular}{l l c c c c c c}
 \toprule
 \multirow{2}{*}{\bf Type} & \multirow{2}{*}{\bf Name}  & \multicolumn{3}{c}{\bf P3 (SQ)}& \multicolumn{3}{c}{\bf P3 (LQ)} \\
  \cmidrule(lr){3-5} \cmidrule(lr){6-8}
 &  & \bf Speech & \bf Pink Noise & \bf Guitar & \bf Speech & \bf Pink Noise & \bf Guitar \\
 \cmidrule(lr){1-8}
\multirow{1}{*}{\bf Conven.}
 & {\bf PESQ} & 0.09 & -0.20 & 0.08 & 0.23 & 0.29 & 0.18 \\
 \cdashline{1-8}
 \multirow{2}{*}{\bf Proposed}
  & {\bf SQ} & \bf 0.76 & \bf 0.74 & \bf 0.40 & - & - &-   \\
  & {\bf LQ} & - & - & - & \bf 0.43 & \bf 0.45 & \bf 0.28 \\
  \cdashline{1-8}
 \multirow{2}{*}{\bf Baseline}
  & {\bf DPLM} & 0.71 & 0.42 & 0.36 & - & - & -  \\
  & {\bf BAMQ} & 0.42 & 0.21 & 0.08 & 0.02 & 0.01 & 0.08  \\

 \bottomrule
\end{tabular}
}

\vspace{-0ex}
\caption{\textbf{Subjective evaluation (2)}: Compared across individual attribute ratings. Models include: Conventional model (e.g., \PESQ), our proposed models (\textit{SQ} and \textit{LQ} models), and \DPLM\ and \BAMQ, as baseline. Spearman Correlation (SC). $\uparrow$ is better.}
\label{table_mos2}
\vspace{-6ex}
\end{table}

\noindent {\bf Monotonicity}
To show our metric's monotonicity with increasing angular distance and noise, we create a test dataset having non-matched audio recordings at increasing perturbation levels. To check for monotonicity in \spatializationquality, we evaluate \OURS\ between a fixed reference, and
a moving test source for three different source positions compared across recordings of \emph{non-matched} content (see Fig~\ref{compare_static_dynamic}). Manocha \etal~\cite{manocha2021dplm} showed the trend for recordings having matched content, and here we see that the general trend is similar.
To check for monotonicity in \listeningquality, we calculate the Spearman rank order correlation (SC) between the distance from the metric and the perturbation levels.
The absolute correlation is high suggesting that the score from the metric decreases with increasing noise level.

\subsection{Subjective evaluations}
\label{sub_eval}

We use previously published diverse third-party studies to verify that our trained metric correlates well with subjective ratings related to their task. We compute the correlation between the model’s predicted distance with the publicly available subjective ratings, using SC. These scores are evaluated per condition where we average scores for each condition. We choose four distinct classes of available datasets for our analysis: i) (P1 and P1\textquotesingle) Bilateral Ambisonics~\cite{ben2021binaural}; ii) (P2) Spherical Microphone Array~\cite{lubeck2020perceptual}; iii) (P3) Headphone Equalization~\cite{engel2019effect}; and iv) (P4) Bitrate Compressed Ambisonics~\cite{rudzki2019perceptual}. All datasets have an overall rating of quality, but P3 additionally has individual ratings for \spatializationquality\ and \listeningquality. 
%
%
For more information on the datasets, please refer to Manocha \etal~\cite{manocha2021dplm}.

   
   
   
\begin{table}[b!]
\vspace{-0.20in}
\centering
\renewcommand{\arraystretch}{1.1}
\resizebox{\columnwidth}{!}{
 \begin{tabular}{l l c c c c c c}
 \toprule
 \multirow{2}{*}{\bf Type} & \multirow{2}{*}{\bf Name}  & \multicolumn{3}{c}{\bf P3 (SQ)}& \multicolumn{3}{c}{\bf P3 (LQ)} \\
  \cmidrule(lr){3-5} \cmidrule(lr){6-8}
 &  & \bf Speech & \bf Pink Noise & \bf Guitar & \bf Speech & \bf Pink Noise & \bf Guitar \\
 \cmidrule(lr){1-8}
 \multirow{2}{*}{\bf Individual M.}
  & {\bf SQ} & 0.75 &  0.61 & 0.21 & - & - &-   \\
  & {\bf LQ} & - & - & - & 0.40 & 0.36 & 0.26 \\
  \cdashline{1-8}
 \multirow{2}{*}{\bf Proposed}
  & {\bf SQ} & \bf 0.76 & \bf 0.74 & \bf 0.40 & - & - &- \\
  & {\bf LQ} & - & - & - & \bf 0.43 & \bf 0.45 & \bf 0.28 \\
 \bottomrule
\end{tabular}
}

\vspace{-0ex}
\caption{\textbf{Ablations}: Models include: Individual trained models for Spatialization and Listening Quality, as well as our Multi-task framework models. Spearman Correlation (SC). $\uparrow$ is better.}
\label{ablation2}
\vspace{-6ex}
\end{table}

Results are displayed in Tables~\ref{table_mos1} and~\ref{table_mos2}. Next, we summarize with a few observations:
\vspace{-0.05in}
\begin{itemize} [leftmargin=0.33cm]

  \setlength{\itemsep}{0pt}
  \setlength{\parskip}{0pt}
  \setlength{\parsep}{0pt}

    \item In Table~\ref{table_mos1} when compared across overall listening subjective ratings, our \overallquality\ metric scores higher correlations than the \spatializationquality\ metric across all datasets suggesting that \overallquality\ implicitly learns useful features combining \spatializationquality\ and \listeningquality\ attributes, showing the usefulness of MTL.
    
    
    \item Our \spatializationquality\ metric obtains higher correlations than \DPLM\ also suggesting the usefulness of our MTL framework, as well as our new \emph{EMD} objective.
    
    \item Our \listeningquality\ metric obtains higher correlations than \PESQ\ which suggests that it considers all channels to predict similarity, rather than evaluating each channel individually like \PESQ.
    
    \item Correlations across P3 are low but have improved slightly compared to \DPLM. This suggests that our model fails to capture subtle differences, some of which are close to JNDs.
    
    \item Overall, our proposed metric \OURS\ achieves the best performance, and shows higher generalizability across a wide variety of attributes, and perturbations.
    
\end{itemize}

\subsection{Ablations}
\label{ablations}
We perform ablation studies to better understand the influence of different components of our metric. We show the advantange of our MTL framework, and also show the influence of different input representations to our model.

\noindent {\bf Multi-task learning}
In this ablation, we assess the significance of the multi-task formulation that predicts \listeningquality\ and \spatializationquality\ together. Our results show that training to predict both together results in higher performance than training the two models individually; often the improvement is of the order of 20 to 30\% (see table~\ref{ablation2}).

\noindent {\bf Different input feature representations}
To evaluate how the metric correlates with different types of input representations, we train another set of models using either magnitude spectrum or phase spectrum, and compare performance with \OURS\ which is based on both magnitude and phase spectrum features. We observe that phase features are better correlated with subjective ratings than magnitude features. We also observe that \OURS\ scores the highest correlations, suggesting that it makes use of both sets of features.

\begin{table}[t!]
\vspace{-3.5ex}
\resizebox{\columnwidth}{!}{ 
 \begin{tabular}{l c c c c c}
 \toprule
 {\bf } &  {\bf PESQ$\uparrow$} & {\bf STOI$\uparrow$} & {\bf L2$\downarrow$} & {\bf M.STFT$\downarrow$} & {\bf Si-SDR$\uparrow$} 
  \\ \toprule
  {\bf Noisy} 
 &  1.15 & 70.9 & 0.058 & 0.32 & -1.51 \\
 \midrule
  {\bf LogMSE} 
 &  1.65 & 83.6 & 0.013 & 0.18 & 9.15 \\
  {\bf Scratch \OURS} 
 &  1.75 &  85.88 &  0.011 & 0.17 &  10.31 \\
 {\bf Finetune \OURS} 
 & \bf 1.83 & \bf 86.50 & \bf 0.008 & \bf 0.13 & \bf 10.9 \\
 \bottomrule
\end{tabular}
}
\caption{{\bf Evaluation of enhancement} models using a held out test set with objective measures.}
\vspace{-6ex}
\label{table_denoising}
\end{table}

\vspace{-0.1in}
\subsection{Binaural speech enhancement}
\label{enhancement}
 

To further demonstrate the effectiveness of our metric, we design a binaural speech enhancement model that is based on a popular U-Net type convolutional neural network~\cite{tan2019learning}. The input to the model is the real and imaginary components of the STFT of the binaural signal, and the outputs are the complex ratio masks.

The baseline enhancement model is trained using Log Mean Square Error (LogMSE) between the estimated and target real and imaginary STFT components for both channels. We supplement \OURS\ in two different ways: (i)~Scratch \OURS\ - where we train using LogMSE and \OURS\ from scratch; and (ii)~Finetune \OURS\ - where we first pretrain on LogMSE and finetune on \OURS.

For evaluation, we randomly select 2000 unseen audio recordings from an unseen dataset~\cite{panayotov2015librispeech,algazi2001cipic}. Following prior work, we evaluate the quality of enhanced binaural recordings using a variety of objective measures: i) PESQ (from 0.5 to 4.5); (ii) Short Time Objective Intelligibility (STOI) (from 0 to 100); (iii) L2 distance on the waveform; (iv) Scale invariant signal to distortion ratio (Si-SDR); and v) Multi-resolution STFT at various FFT, hop, and window sizes, all evaluated over each channel separately, and then averaged to get one value.

As we see in Table~\ref{table_denoising}, both of our models perform better than the baseline. We observe that our Finetune \OURS\ model scores the best objective scores. This highlights the usefulness of using \OURS\ in audio similarity tasks, especially in identifying and eliminating minor human perceptible artifacts that are not captured by traditional losses. 
\section{Conclusions and future work}

We present \OURS, a general purpose, differentiable, non-matching, non-clean-reference based objective metric to assess listening quality and spatial localization differences between \emph{any} two binaural signals. We show the utility of MTL and deep-feature distances that guide the model to correlate well with human subjective ratings without any perceptual training or calibration. In the future, we would like to include more spatial quality attributes like perception of distance and coloration.

\clearpage
\bibliographystyle{IEEEtran}

\bibliography{mybib}

\begin{thebibliography}{10}
\providecommand{\url}[1]{#1}
\csname url@samestyle\endcsname
\providecommand{\newblock}{\relax}
\providecommand{\bibinfo}[2]{#2}
\providecommand{\BIBentrySTDinterwordspacing}{\spaceskip=0pt\relax}
\providecommand{\BIBentryALTinterwordstretchfactor}{4}
\providecommand{\BIBentryALTinterwordspacing}{\spaceskip=\fontdimen2\font plus
\BIBentryALTinterwordstretchfactor\fontdimen3\font minus
  \fontdimen4\font\relax}
\providecommand{\BIBforeignlanguage}[2]{{%
\expandafter\ifx\csname l@#1\endcsname\relax
\typeout{** WARNING: IEEEtran.bst: No hyphenation pattern has been}%
\typeout{** loaded for the language `#1'. Using the pattern for}%
\typeout{** the default language instead.}%
\else
\language=\csname l@#1\endcsname
\fi
#2}}
\providecommand{\BIBdecl}{\relax}
\BIBdecl

\bibitem{werner2014influence}
S.~Werner and F.~Klein, ``Influence of context dependent quality parameters on
  the perception of externalization and direction of an auditory event,'' in
  \emph{AES Conference: 55th International Conference: Spatial Audio}.\hskip
  1em plus 0.5em minus 0.4em\relax AES, 2014.

\bibitem{lindau2014spatial}
A.~Lindau, ``Spatial audio quality inventory ({SAQI}). test manual.''

\bibitem{kampf2010standardization}
S.~K{\"a}mpf, J.~Liebetrau, S.~Schneider \emph{et~al.}, ``Standardization of
  {PEAQ-MC}: Extension of {ITU-R} bs. 1387-1 to multichannel audio,'' in
  \emph{AES Conference: 40th International Conference: Spatial Audio: Sense the
  Sound of Space}.\hskip 1em plus 0.5em minus 0.4em\relax AES, 2010.

\bibitem{flessner2017assessment}
J.-H. Fle{\ss}ner, R.~Huber, and S.~D. Ewert, ``Assessment and prediction of
  binaural aspects of audio quality,'' \emph{Journal of the AES}, vol.~65,
  no.~11, pp. 929--942, 2017.

\bibitem{seo2013perceptual}
J.-H. Seo, S.~B. Chon, K.-M. Sung \emph{et~al.}, ``Perceptual objective quality
  evaluation method for high quality multichannel audio codecs,'' \emph{Journal
  of the AES}, vol.~61, no. 7/8, pp. 535--545.

\bibitem{takanen2012binaural}
M.~Takanen and G.~Lorho, ``A binaural auditory model for the evaluation of
  reproduced stereophonic sound,'' in \emph{AES Conference: 45th International
  Conference: Applications of Time-Frequency Processing in Audio}.\hskip 1em
  plus 0.5em minus 0.4em\relax AES, 2012.

\bibitem{narbutt2020ambiqual}
M.~Narbutt, J.~Skoglund, A.~Allen \emph{et~al.}, ``{AMBIQUAL}: Towards a
  quality metric for headphone rendered compressed ambisonic spatial audio,''
  \emph{Applied Sciences}, vol.~10, no.~9, 2020.

\bibitem{manocha2021dplm}
P.~Manocha, A.~Kumar, B.~Xu \emph{et~al.}, ``{DPLM}: A deep perceptual
  spatial-audio localization metric,'' in \emph{WASPAA}.\hskip 1em plus 0.5em
  minus 0.4em\relax IEEE, 2021, pp. 6--10.

\bibitem{manocha2021noresqa}
P.~Manocha, B.~Xu, and A.~Kumar, ``{NORESQA}: A framework for speech quality
  assessment using non-matching references,'' \emph{NeurIPS}, vol.~34, 2021.

\bibitem{wang2014learning}
J.~Wang, Y.~Song, T.~Leung \emph{et~al.}, ``Learning fine-grained image
  similarity with deep ranking,'' in \emph{Proceedings of the IEEE conference
  on computer vision and pattern recognition}, 2014, pp. 1386--1393.

\bibitem{hoffer2015deep}
E.~Hoffer and N.~Ailon, ``Deep metric learning using triplet network,'' in
  \emph{International workshop on similarity-based pattern recognition}.\hskip
  1em plus 0.5em minus 0.4em\relax Springer, 2015, pp. 84--92.

\bibitem{caruana1997multitask}
R.~Caruana, ``Multitask learning,'' \emph{Machine learning}, vol.~28, no.~1,
  pp. 41--75, 1997.

\bibitem{fu2016snr}
S.-W. Fu, Y.~Tsao, and X.~Lu, ``{SNR-Aware} convolutional neural network
  modeling for speech enhancement.'' in \emph{Interspeech}, 2016, pp.
  3768--3772.

\bibitem{chen2015multitask}
D.~Chen and B.~K.-W. Mak, ``Multitask learning of deep neural networks for
  low-resource speech recognition,'' \emph{IEEE/ACM TASLP}, vol.~23, no.~7, pp.
  1172--1183, 2015.

\bibitem{szegedy2015going}
C.~Szegedy, W.~Liu, Y.~Jia \emph{et~al.}, ``Going deeper with convolutions,''
  in \emph{Proceedings of the IEEE Conference on Computer Vision and Pattern
  Recognition}, 2015.

\bibitem{salimans2016weight}
T.~Salimans and D.~P. Kingma, ``Weight normalization: A simple
  reparameterization to accelerate training of deep neural networks,''
  \emph{Advances in neural information processing systems}, vol.~29, 2016.

\bibitem{manocha2020differentiable}
P.~Manocha, A.~Finkelstein, R.~Zhang \emph{et~al.}, ``A differentiable
  perceptual audio metric learned from just noticeable differences,''
  \emph{Interspeech}, 2020.

\bibitem{garofolo1993darpa}
J.~S. Garofolo, L.~F. Lamel, W.~M. Fisher \emph{et~al.}, ``{DARPA TIMIT}
  acoustic-phonetic continous speech corpus cd-rom. nist speech disc 1-1.1,''
  \emph{NASA STI/Recon technical report n}, vol.~93, p. 27403, 1993.

\bibitem{winter2016database}
F.~Winter, H.~Wierstorf, A.~Podlubne \emph{et~al.}, ``Database of binaural room
  impulse responses of an apartment-like environment,'' in \emph{AES Convention
  140}.\hskip 1em plus 0.5em minus 0.4em\relax AES, 2016.

\bibitem{jeub2009binaural}
M.~Jeub, M.~Schafer, and P.~Vary, ``A binaural room impulse response database
  for the evaluation of dereverberation algorithms,'' in \emph{2009 16th
  International Conference on Digital Signal Processing}.\hskip 1em plus 0.5em
  minus 0.4em\relax IEEE, 2009, pp. 1--5.

\bibitem{aspock2020bras}
L.~Asp{\"o}ck, F.~Brinkmann, D.~Ackermann \emph{et~al.}, ``{BRAS}-{Benchmark}
  for room acoustical simulation,'' 2020.

\bibitem{bacila2019360}
B.~I. Bacila and H.~Lee, ``360° binaural room impulse response ({BRIR})
  database for {6DOF} spatial perception research,'' in \emph{AES Convention
  146}.\hskip 1em plus 0.5em minus 0.4em\relax AES, 2019.

\bibitem{mittag_christina_2016_206860}
\BIBentryALTinterwordspacing
C.~Mittag, M.~Böhme, and S.~Werner, ``{Dataset of KEMAR-BRIRs measured at
  several positions and head orientations in a real room},'' Dec. 2016.
  [Online]. Available: \url{https://doi.org/10.5281/zenodo.206860}
\BIBentrySTDinterwordspacing

\bibitem{francombe2016iosr}
J.~Francombe, ``Iosr listening room multichannel {BRIR} dataset.''

\bibitem{kayser2009database}
H.~Kayser, S.~D. Ewert, J.~Anem{\"u}ller \emph{et~al.}, ``Database of
  multichannel in-ear and behind-the-ear head-related and binaural room impulse
  responses,'' \emph{EURASIP}, vol. 2009, pp. 1--10, 2009.

\bibitem{erbes2015database}
V.~Erbes, M.~Geier, S.~Weinzierl \emph{et~al.}, ``Database of single-channel
  and binaural room impulse responses of a 64-channel loudspeaker array,'' in
  \emph{AES Convention 138}.\hskip 1em plus 0.5em minus 0.4em\relax AES, 2015.

\bibitem{wierstorf2011free}
H.~Wierstorf, M.~Geier, and S.~Spors, ``A free database of head related impulse
  response measurements in the horizontal plane with multiple distances,'' in
  \emph{AES Convention 130}.\hskip 1em plus 0.5em minus 0.4em\relax AES, 2011.

\bibitem{satongar2014measurement}
D.~Satongar, Y.~W. Lam, and C.~Pike, ``Measurement and analysis of a spatially
  sampled binaural room impulse response dataset,'' in \emph{21st International
  Congress on Sound and Vibration}, 2014.

\bibitem{reddy2020interspeech}
C.~K. Reddy, E.~Beyrami, H.~Dubey \emph{et~al.}, ``The interspeech 2020 deep
  noise suppression challenge: Datasets, subjective speech quality and testing
  framework,'' \emph{arXiv arXiv:2001.08662}, 2020.

\bibitem{foster2015chime}
P.~Foster, S.~Sigtia, S.~Krstulovic \emph{et~al.}, ``Chime: A dataset for sound
  source recognition in a domestic environment,'' in \emph{IEEE WASPAA}, 2015.

\bibitem{weisser2019ambisonic}
A.~Weisser, J.~M. Buchholz, C.~Oreinos \emph{et~al.}, ``The ambisonic
  recordings of typical environments ({ARTE}) database,'' \emph{Acta Acustica
  United With Acustica}, vol. 105, no.~4, 2019.

\bibitem{cornelis2009theoretical}
B.~Cornelis, S.~Doclo, T.~Van~dan Bogaert \emph{et~al.}, ``Theoretical analysis
  of binaural multimicrophone noise reduction techniques,'' \emph{IEEE TASLP},
  vol.~18, no.~2, pp. 342--355, 2009.

\bibitem{harwood2017smart}
B.~Harwood, V.~Kumar~BG, G.~Carneiro \emph{et~al.}, ``Smart mining for deep
  metric learning,'' in \emph{CVPR}, 2017, pp. 2821--2829.

\bibitem{rix2001perceptual}
A.~W. Rix, J.~G. Beerends, M.~P. Hollier \emph{et~al.}, ``Perceptual evaluation
  of speech quality ({PESQ})-a new method for speech quality assessment of
  telephone networks and codecs,'' in \emph{2001 IEEE ICASSP}, vol.~2.\hskip
  1em plus 0.5em minus 0.4em\relax IEEE, 2001, pp. 749--752.

\bibitem{ben2021binaural}
Z.~Ben-Hur, D.~L. Alon, R.~Mehra \emph{et~al.}, ``Binaural reproduction based
  on bilateral ambisonics and ear-aligned hrtfs,'' \emph{IEEE/ACM TASLP},
  vol.~29, pp. 901--913, 2021.

\bibitem{lubeck2020perceptual}
T.~L{\"u}beck, H.~Helmholz, J.~M. Arend \emph{et~al.}, ``Perceptual evaluation
  of mitigation approaches of impairments due to spatial undersampling in
  binaural rendering of spherical microphone array data,'' \emph{Journal of the
  AES}, vol.~68, no.~6, pp. 428--440, 2020.

\bibitem{engel2019effect}
I.~Engel, D.~L. Alon, P.~W. Robinson \emph{et~al.}, ``The effect of generic
  headphone compensation on binaural renderings,'' in \emph{AES Conference:
  2019 AES International Conference on Immersive and Interactive Audio}.\hskip
  1em plus 0.5em minus 0.4em\relax AES, 2019.

\bibitem{rudzki2019perceptual}
T.~Rudzki, I.~Gomez-Lanzaco, P.~Hening \emph{et~al.}, ``Perceptual evaluation
  of bitrate compressed ambisonic scenes in loudspeaker based reproduction,''
  in \emph{AES International Conference on Immersive and Interactive Audio},
  2019.

\bibitem{tan2019learning}
K.~Tan and D.~Wang, ``Learning complex spectral mapping with gated
  convolutional recurrent networks for monaural speech enhancement,''
  \emph{IEEE/ACM TASLP}, vol.~28, pp. 380--390, 2019.

\bibitem{panayotov2015librispeech}
V.~Panayotov, G.~Chen, D.~Povey \emph{et~al.}, ``Librispeech: an asr corpus
  based on public domain audio books,'' in \emph{ICASSP}.\hskip 1em plus 0.5em
  minus 0.4em\relax IEEE, 2015, pp. 5206--5210.

\bibitem{algazi2001cipic}
V.~R. Algazi, R.~O. Duda, D.~M. Thompson \emph{et~al.}, ``The {CIPIC} hrtf
  database,'' in \emph{Proceedings of the 2001 IEEE Workshop on the
  Applications of Signal Processing to Audio and Acoustics (Cat. No.
  01TH8575)}.\hskip 1em plus 0.5em minus 0.4em\relax IEEE, 2001, pp. 99--102.

\end{thebibliography}

\end{document}